\documentclass[%
reprint,
superscriptaddress,
amsmath,amssymb,
aps,
pra,
]{revtex4-2}
\usepackage{graphicx,amsmath,amssymb,amsfonts,dsfont,subfigure,color}
\usepackage{bm,txfonts}
\usepackage{verbatim}
\usepackage{dcolumn}% Align table columns on decimal point
\usepackage{bm}% bold math
\usepackage{epsf}
\usepackage{xcolor}
\usepackage{hyperref}
\usepackage{hhline}
\usepackage{float}
\usepackage{enumerate}
\usepackage{bbm}
\usepackage{lipsum}
\usepackage{mathrsfs}
\usepackage[normalem]{ulem}
\begin{document}
	
	\preprint{APS/123-QED}
	\title[Short Title]{Optimizing Four-Wave Mixing in Rydberg Atoms for Microwave-Optical Conversion}
\author{Ning Ji}
\author{Yanzhao Liang}
\author{Wanrang Yu}
\author{Qiuyu Yin}
%\affiliation{University, Guangdong , China}
\author{Thibault Vogt}
\email{ttvogt@mail.sysu.edu.cn}
\affiliation{School of Physics and Astronomy, Sun Yat-sen University, Zhuhai 519082, China}
\begin{abstract}
We perform a numerical and analytical investigation of microwave-to-optical conversion based on four-wave mixing in Rydberg atoms. Our work demonstrates that both all-resonant and off-resonant frequency-mixing configurations achieve near-unit photon conversion efficiencies. We review the conditions that can lead to the presence of two possible dark states. We find that for both configurations, one of the dark states can be detrimental at high microwave powers, and show that an additional limitation to all-resonant frequency mixing is microwave-induced fluorescence. Finally, we confirm that the off-resonant configuration is more appropriate as it allows for efficient photon conversion on a wider range of input microwave intensities with reduced total power of the auxiliary fields. 
\end{abstract}
\maketitle
\section{Introduction}
\indent Efficient interconversion between microwaves and optical fields is of strategic importance for interconnecting superconducting quantum processors and linking them to other quantum platforms~\cite{2020Lambert,2021Han}. Indeed, superconducting qubit systems work in the microwave range of frequencies but microwave photons cannot be used for transporting quantum information at room temperature over long distances because of thermal noise~\cite{2018Kurpiers}. Microwave-optical quantum conversion can solve this problem by enabling long-distance quantum communication between these systems using optical photons.\\
\indent One promising approach to microwave-optical transduction relies on frequency mixing in a cold atom ensemble where the microwave field is strongly coupled to an atomic transition between two Rydberg energy levels~\cite{2016Kiffner,2019Petrosyan,2018Coherent,2019Efficient,2022High,2023Continuous,2023Kumar}. The Rydberg atom approach offers large conversion bandwidth matching well that of single photon sources used for quantum communications~\cite{2016Rambach}. Both four and six-wave mixing have already been employed to achieve high conversion efficiencies~\cite{2018Coherent,2019Efficient,2022High,2023Continuous,2023Kumar}. Being able to obtain at the same time large efficiency and bandwidth makes frequency mixing via Rydberg energy levels particularly advantageous in comparison to other technologies~\cite{2018Fan,2018Higginbotham,2020Mirhosseini,2022Delaney}. In addition, recent technological advances in hybrid systems combining Rydberg atoms and superconducting qubits bring great promise to this approach~\cite{2015Teixeira,2020Morgan,2022Kaiser,2023Kumar}.\\
\indent The earliest demonstrations of microwave-to-optical conversion via frequency mixing in Rydberg atom systems employed all-resonant fields, in contrast to recent achievements~\cite{2022High,2023Continuous,2023Kumar}. One important problem of all-resonant wave-mixing that has been noticed is the existence of a dark state~\cite{2018Coherent,2022High}. The atoms get trapped in the dark state that is decoupled from the inter-converted microwave and optical fields, eventually limiting the efficiency of the conversion. However, the existence of a dark state was not perceived as a major issue in several other studies of near-resonant frequency-mixing~\cite{2010Jen,2010Gogyan,2019Efficient}. Moreover, trapping in the dark state occurs if certain conditions on the fields intensities are satisfied \textcolor{blue}{~\cite{2019Classification}}, and these conditions are not necessarily incompatible with detuning some of the fields in the system.\\
\indent In this paper, we propose a comprehensive study of microwave-to-optical conversion based on four wave mixing in cold atoms. In section~\ref{model}, we present the formalism and derive the relevant analytical solutions of Maxwell-Bloch equations in steady state. We propose a complete review of the conditions that lead to the presence of two possible dark states in the system. In section~\ref{result}, we perform a thorough comparison of full-numerical results with analytical solutions for both on and off-resonant configurations. In the on-resonance case, near-unit photon conversion efficiency is easily achieved at low microwave intensity.  However, microwave-induced fluorescence, in combination with the presence of one of the dark states, limits the dynamical range of the transducer.  We explain how detuning two of the fields in a two-photon excitation configuration via off-resonant intermediate state is employed to circumvent microwave-induced fluorescence. In this case, the dynamical range is only limited by the dark state. Interestingly for quantum applications, off-resonant frequency mixing allows for using less power of the auxiliary optical fields. One drawback of the off-resonance configuration is its slightly reduced bandwidth. Finally, a full-numerical optimization is carried out, whose results show that efficiencies larger than 95\% can be reached with both configurations.
\section{Formalism}
\label{model}
In this section, we first present the frequency mixing scheme consisting of four fields interacting near-resonantly with four-level atoms in an atomic cloud. Next, the formalism is provided, based on the steady-state solution of Maxwell-Bloch equations. Although the model is valid for bidirectional microwave-optical conversion, for simplicity, only the microwave-to-optical scenario is being investigated. Finally, we address the issue of the dark states.\\ 
\indent As illustrated in figure \ref{f1}(a), a pump field P of Rabi frequency $\Omega_{P}$ drives the transition from the atomic ground state $\left | 1  \right \rangle$ to the Rydberg state $\left | 2  \right \rangle$. The microwave field M to be converted, of Rabi frequency $\Omega_{M}$, drives the transition from the Rydberg state $\left | 2  \right \rangle$ to Rydberg state $\left | 3  \right \rangle$. A strong control field of Rabi frequency $\Omega_{C}$ acts on the transition between the Rydberg state $\left | 3  \right \rangle$ and the low-lying excited state $\left | 4  \right \rangle$. Once the auxiliary fields P and C and microwave field M overlap inside the atomic cloud, the coherence induced between the ground state $\left | 1  \right \rangle$ and the intermediate state $\left | 4  \right \rangle$ triggers the generation of the converted optical field L with frequency $\nu_{L}=\nu_{P}+\nu_{M}-\nu_{C}$ (where $\nu_{X}$ is the frequency of field X, with $\textrm{X}\in\{P, M, C,L\}$). The fields propagate collinearly inside the atomic cloud as sketched in Fig.~\ref{f1}(c) and they are assumed to fulfill the phase matching condition $\textbf{k}_{L}=\textbf{k}_{P}+\textbf{k}_{M}-\textbf{k}_{C}$, where $\textbf{k}_{X}$ represents the wave vector of the corresponding field. 
	%The wave vectors of the microwave field $\textbf{k}_{M}$ is negligible, since they are much smaller than thoes of the optical fields.
In the framework of the electric dipole and rotating-wave approximations, the Hamiltonian governing the interaction of a four-level atom with external fields is given as follows:	
\begin{figure}[t]
	\centering
	\includegraphics[width=0.9\linewidth]{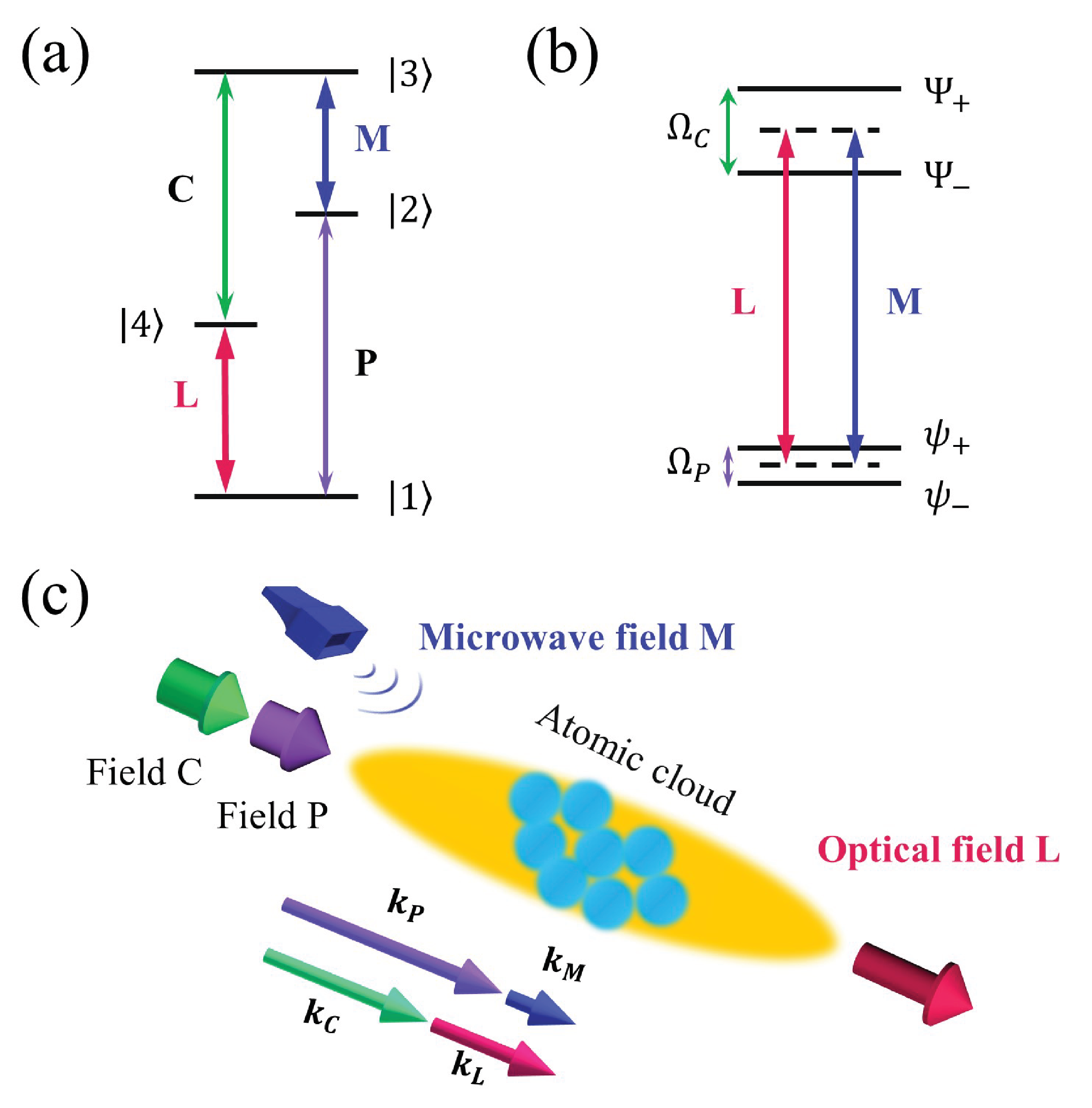}
	\caption{Schematic diagrams of four-wave mixing: (a)  symmetric ladder excitation configuration; (b) dressed states of the atom interacting with two strong auxiliary fields; (c) geometry of the setup enabling microwave-to-optical conversion. Frequency mixing is performed in a Gaussian-distributed atomic sample. The laser beams are focused on the atomic ensemble, whereas the microwave field is emitted from a horn antenna.}
	\label{f1}
\end{figure}

\begin{eqnarray}
\label{eq1}
H= &-&\hbar\left[\Delta_{P}A_{22}+(\Delta_{P}+\Delta_{M})A_{33}+(\Delta_{P}+\Delta_{M}-\Delta_{C})A_{44}\right] \nonumber\\
&+&\frac{\hbar}{2}\left(\Omega_{P}A_{21}+\Omega_{M}A_{32}+\Omega_{C}A_{34 }+\Omega_{L}A_{41}+\mathrm{H.c.}\right),
\end{eqnarray}
where $A_{ji}=|j\rangle\langle i| $, and $\Delta_{X}$ represents the detuning of field $X$ that drives the corresponding transition $\left | i  \right \rangle $$\rightarrow$$\left | j  \right \rangle $ with dipole matrix element $d_{ji}$ and Rabi frequency $\Omega_{X}=-\varepsilon_{X}d_{ji}/\hbar$.  The atomic coherences are given by the steady-state solution $\rho$ of the following Markovian master equation:
\begin{eqnarray}
\label{eq2}	
\partial_t\rho=-\frac{i}{\hbar}[H,\rho]+\mathcal{L}_\Gamma\rho+\mathcal{L}_D \rho.
\end{eqnarray}	
In this equation, $\mathcal{L}_\Gamma\rho$ accounts for the effect of spontaneous emission, and is written as 
\begin{eqnarray}
\label{eq3}	
\mathcal{L}_{\Gamma}\rho=&-&\frac{\gamma_2}{2}\left(A_{22}\rho+\rho A_{22}-2A_{12}\rho A_{21}\right)  \nonumber\\
&-&\frac{\gamma_3}{2}\left(A_{33}\rho+\rho A_{33}-2A_{23}\rho A_{32}\right) \nonumber\\
&-&\frac{\Gamma}{2}\left(A_{44}\rho+\rho A_{44}-2A_{14}\rho A_{41}\right),
\end{eqnarray}
where the Lindblad terms proportional to $\gamma_2$ and $\gamma_3$ correspond to spontaneous decay from the Rydberg states $\left | 2  \right \rangle$ and $\left | 3  \right \rangle$, respectively, and $\Gamma$ is the decay rate from the intermediate state $\left | 4  \right \rangle$ which is about three orders of magnitude larger than  $\gamma_2$ and $\gamma_3$. For simplicity, we assume that $\gamma_2=\gamma_3=\gamma$.
In the numerical simulations, we add a Liouvillian mimicking the effect of the P and C laser phase noises
\begin{eqnarray}
	\label{eq3b}	
	\mathcal{L}_{D}\rho=&-&\frac{\gamma_P}{2}\left(A_{22}\rho+\rho A_{22}-2A_{22}\rho A_{22}\right)  \nonumber\\
	&-&\frac{\gamma_C}{2}\left(A_{33}\rho+\rho A_{33}-2A_{33}\rho A_{33}\right),
\end{eqnarray}
where $\gamma_P$ and $\gamma_C$ are the two relevant dephasing rates. Next, we restrict the problem to the calculation of Maxwell's equations for the M and L fields in the slowly varying envelope approximation, and in steady state. Note that because $\gamma\ll\Gamma$, the steady state is reached for atom-fields interaction times longer than $1/\gamma$ in the limit of weak M and L fields. With these assumptions and neglecting transverse terms that account for diffraction and lensing effects, the fields' amplitudes satisfy the following one-dimensional differential equations:
\begin{subequations}
	\begin{eqnarray}
	\label{eq5a}
	\frac{\partial \Omega_{M}}{\partial z}=i \eta_{M}\rho_{32}, \\
	\label{eq5b}
	\frac{\partial \Omega_{L}}{\partial z}= i \eta_{L} \rho_{41},
	\end{eqnarray}
\end{subequations}
where $\eta_{M (L)}=n_{at}|d_{ji}|^{2}\nu_{M (L)}/(h\epsilon_{0}c)$ is a coupling constant depending on the atomic density $n_{at}$, the free-space permittivity $\epsilon_{0}$, and the speed of light in vacuum $c$, while  $\rho_{41}$ and $\rho_{32}$ are atomic coherences. 
The absorption rates of the P and C fields are negligible because the dipole moments of the $\left | 1  \right \rangle $$\leftrightarrow $$\left | 2  \right \rangle $ and $\left | 4  \right \rangle $$\leftrightarrow $$\left | 3  \right \rangle $ transitions are several orders of magnitude smaller than that of the $\left | 1  \right \rangle $$\leftrightarrow $$\left | 4  \right \rangle $ transition, hence $\Omega_{P}$ and $\Omega_{C}$ keep constant in a very good approximation. 

From the results of Eqs. \eqref{eq5a} and \eqref{eq5b}, the photon conversion efficiency of
the process is calculated as
\begin{eqnarray}
\label{eq7}
\eta=\frac{|\Omega_{L}^{out}|^{2}/\eta_{L}}{|\Omega_{M0}|^{2}/\eta_{M}},
\end{eqnarray}
where $\Omega^{out}_{L}$ is the Rabi frequency of the converted field L emerging from the atomic medium, and $\Omega_{M0}$ is the input microwave Rabi frequency. The conditions providing maximum conversion efficiency are inferred from this formula. 

To provide insight into the photon conversion process, Eq. \eqref{eq2} can be solved using perturbation theory. We obtain the analytical steady-state solution of  Eq. \eqref{eq2} by expansion of $\rho$ in the weak fields $ \Omega_{M}$ and $\Omega_{L}$. To leading order, the coherences $\rho_{32}$ and $\rho_{41}$ are expressed as follows (see Appendix A for more details):
\begin{eqnarray}
	\label{eq4}
\begin{bmatrix}
\rho_{32}\\
\rho_{41}
\end{bmatrix}=
\begin{bmatrix}
\alpha_M&\beta_L \\
\beta_M&\alpha_L
\end{bmatrix}
\begin{bmatrix}
\Omega_M\\
\Omega_L
\end{bmatrix},
\end{eqnarray}
where $\alpha_{M}$ and $\alpha_{L}$ are the self-coupling coefficients induced on the $\left | 2  \right \rangle $$\leftrightarrow $$\left | 3  \right \rangle $ and $\left | 1  \right \rangle $$\leftrightarrow $$\left | 4  \right \rangle $ transitions by the microwave and optical fields, respectively, while $\beta_{L}$ and $\beta_{M}$ are parametric coefficients. The optical field creates a coherence proportional to the parametric coefficient $\beta_{L}$ on the $\left | 2  \right \rangle $$\leftrightarrow $$\left | 3  \right \rangle $ transition, and the microwave field induces a coherence $\propto\beta_{M}$ on the  $\left | 1  \right \rangle $$\leftrightarrow $$\left | 4  \right \rangle $ transition. These four quantities are defined in Appendix A.
Substituting Eq. (\ref{eq4}) into Eqs. (\ref{eq5a})-\eqref{eq5b}, we can obtain analytical expressions for the L and M fields at the position $z$ within the atomic medium as follows:
\begin{subequations}
\begin{eqnarray}
\label{eq6a}
\Omega_{L}(z)&=&\Omega_{M0}a_2\left( e^{i a_1 z}-e^{-i a_1 z}\right)e^{ia_3z},\\
\label{eq6b}
\Omega_{M}(z)&=&\Omega_{M0}\left(a_4e^{i a_1 z}-(a_4-1)e^{-i a_1 z}\right)e^{ia_3z},
\end{eqnarray}
\end{subequations}
where $a_1=\sqrt{\alpha_{L}^{2}\eta_{L}^{2}+\alpha_{M}^{2}\eta_{M}^{2}+2\eta_{L}\eta_{M}(2\beta_{M}\beta_{L}-\alpha_{M}\alpha_{L})}/2$, $a_2=\beta_{M}\eta_{L}/2 a_1$, $a_3=(\alpha_{L}\eta_{L}+\alpha_{M}\eta_{M})/2$, and $a_4=(\alpha_{M}\eta_{M}-\alpha_{L}\eta_{L})/4a_1$ are complex quantities.
Equations \eqref{eq6a} and \eqref{eq6b} are valid considering an atomic cloud with uniform atomic density. They are still valid for a cloud with Gaussian density distribution given that $n_{at}$ represents the peak atomic density and the propagation distance $z$ is replaced with $\tilde{z}$ defined as 
\begin{eqnarray}
\label{eq6c}
\tilde{z}= \int_{-\infty }^{z} e^{-\frac{2 z^{\prime 2}}{w_z^2}}dz^{\prime}.
\end{eqnarray}
Note that the analytical expressions of $\rho_{32}$ and $\rho_{41}$ are greatly simplified in the case where $\Omega_{P}$ is also considered as a weak field for the calculation of $\rho$ from perturbation theory. Applying perturbation theory with weak P field is valid for $\Omega_P<\gamma$, and in this case, the coefficients of the matrix in Eq.~\eqref{eq4} verify $\alpha_{L}=\alpha_{M}=0$, and $\beta_{M}=\beta_{L}^*$, as shown in Appendix A.

The presence of approximate dark states may potentially limit the efficiency. 
For example, when $\Delta_{P}+\Delta_{M}=0$, and if the following condition is satisfied
%nately
\begin{eqnarray}
	\label{eq8}
	\frac{\Omega_{L}\Omega_{M}^{*}}{\Omega_{P}\Omega_{C}^{*}}=1,
\end{eqnarray}
one of the eigenstates of Hamiltonian $H$ is an approximate dark state of the form
\begin{eqnarray}
	\label{eq9}
	\left | D  \right \rangle=\frac{1}{\sqrt{|\Omega_{P}|^{2}+|\Omega_{M}|^{2}}}(\Omega_{M}^{*}\left | 1  \right \rangle-\Omega_{P}\left | 3  \right \rangle).
\end{eqnarray}
This state exhibits nonzero population only in long-lived states $ | 1  \rangle$ and $ | 3  \rangle$. The atomic medium is decoupled from all the fields when all the atoms are trapped in the $ | D  \rangle$ state. We stress here that $ | D  \rangle$ and the condition in Eq. \eqref{eq8} are independent of $\Delta_C$ and $\Delta_P$, provided $\Delta_{P}+\Delta_{M}=0$. This dark state poses a problem mainly for large $M$ and $L$ fields intensities.
Definitely, detuning some of the fields does not preclude trapping in the dark state. However, this remains highly beneficial for increasing the dynamical range of the converter as discussed in Section \ref{result}. 

More generally, given the projector $P$ on the subspace spanned by the three long-lived states $ |1\rangle$, $|2\rangle$, and $|3 \rangle $, one eigenstate $| \psi_k \rangle$ of the operator $P H P$ is a dark state if the two following conditions are met. First, the equality $[H,\rho_k] = 0$ must be satisfied,  where $ \rho_k=|\psi_k\rangle \langle\psi_k|$, which upon multiplication on the left by $\langle 4|$ leads to $\langle 4|H|\psi_k\rangle = 0 $~\cite{2019Classification}. Second, we should have $\langle 3| \rho_k| 2\rangle =0$ for the field $M$ to be decoupled from the atoms during its propagation. When $\Delta_{P}+\Delta_{M}=0$, the state in Eq. \eqref{eq9} is the sole possible pure state satisfying both conditions.
If several eigenstates $|\psi_k\rangle$ satisfy $\langle 4|H|\psi_k\rangle = 0 $, a dark mixed state $\rho=\sum p_k |\psi_k\rangle \langle\psi_k|$ can exist, given $p_k\geq 0$ and $\sum p_k=1$, and as long as $\langle 3| \rho| 2\rangle =0$~\cite{2019Classification}. For $\Delta_{P}+\Delta_{M}=0$, such an additional dark state is present when (see Appendix B)
\begin{eqnarray}
	\label{eq9b}
	\Omega_{L}\Omega_{P}^{*}=-\Omega_{M}\Omega_{C}^{*}.
\end{eqnarray}
This second dark state could be very detrimental as it would limit the efficiency $\propto \left| \Omega_{L}/\Omega_{M} \right|^2$ even for low power of the field to be converted. In practice, trapping in this dark state barely occurs due to the presence of the non-zero decay rate $\gamma$ which was neglected while assuming that $ |1\rangle$, $|2\rangle$, and $|3 \rangle $ are long-lived states. There is some evidence that this dark state can occur when $\Omega_{P} \sim \Omega_{C}\gg \gamma$ as explained in Appendix B. However, we are not interested in this regime of excitation, in which the dressing by the P field as sketched in Fig. \ref{f1}(b) causes large scattering, and EIT on the $|1\rangle\rightarrow |4\rangle \rightarrow |3\rangle$ ladder-scheme becomes inefficient.

\section{Results And Discussions}
\label{result}
\subsection{Example system}
\label{expamle}
After having reviewed the formalism, we come to analyze practical configurations for experiments. We commence this section with a discussion of one possible realization of our scheme based on a Gaussian-distributed atomic ensemble of cesium atoms of $1/e^2$ radius $w_z=1$~mm. The configuration is such that $\left | 1  \right \rangle \equiv\left | 6S_{1/2}, F=4, M_F=4  \right \rangle$, $\left | 2  \right \rangle \equiv\left | 43P_{3/2}, M_J=3/2  \right \rangle$, $\left | 3  \right \rangle \equiv\left | 42D_{5/2}, M_J=5/2  \right \rangle$, and $\left | 4  \right \rangle \equiv\left | 6P_{3/2}, F=5, M_F=5  \right \rangle$
, with transition dipole moments $d_{12}=0.0033$~a.u., $d_{14}=3.17$~a.u., $d_{23}=1468.43$~a.u., and $d_{43}=0.029$~a.u. The decay rate of state $|4\rangle$ is $\Gamma=2\pi \times 5.22$~MHz. We work at cryogenic temperatures $<4$~K, where the effect of blackbody radiation on the  lifetime of the Rydberg states is negligible.
Consistently with our analytical calculations, we choose a unique decay rate $\gamma=2 \pi \times 10$~kHz that compares well with the decay rates of the chosen Rydberg states, and $\gamma_C = \gamma_P = \gamma_d = 2 \pi \times 1$~kHz. The computations are performed for peak atomic densities $n_{at}$ of the atomic cloud given the corresponding optical depths (OD)  for the $L$ field of wavelength $\lambda_L\approx 852.3$~nm, $\rm{OD}= (8 \pi)^{-1/2}$$3\lambda_{L}^{2}w_{z}n_{at}$. Most results are presented for a very low input microwave Rabi frequency of $\Omega_{M0}=2\pi\times0.01$ MHz which corresponds to about 10 microwave photons in a volume of the order of $(c/\nu_M)^3$, where $\nu_M\approx 9.9$~GHz. Controlling such a weak field is difficult in practice. It was chosen for avoiding issues related to the presence of the dark states, and the reported results are still valid for a microwave field at least one order of magnitude stronger on the full range of input parameters being explored in this paper.

\begin{figure}[htbp]
	\centering
	\includegraphics[width=1.0\linewidth]{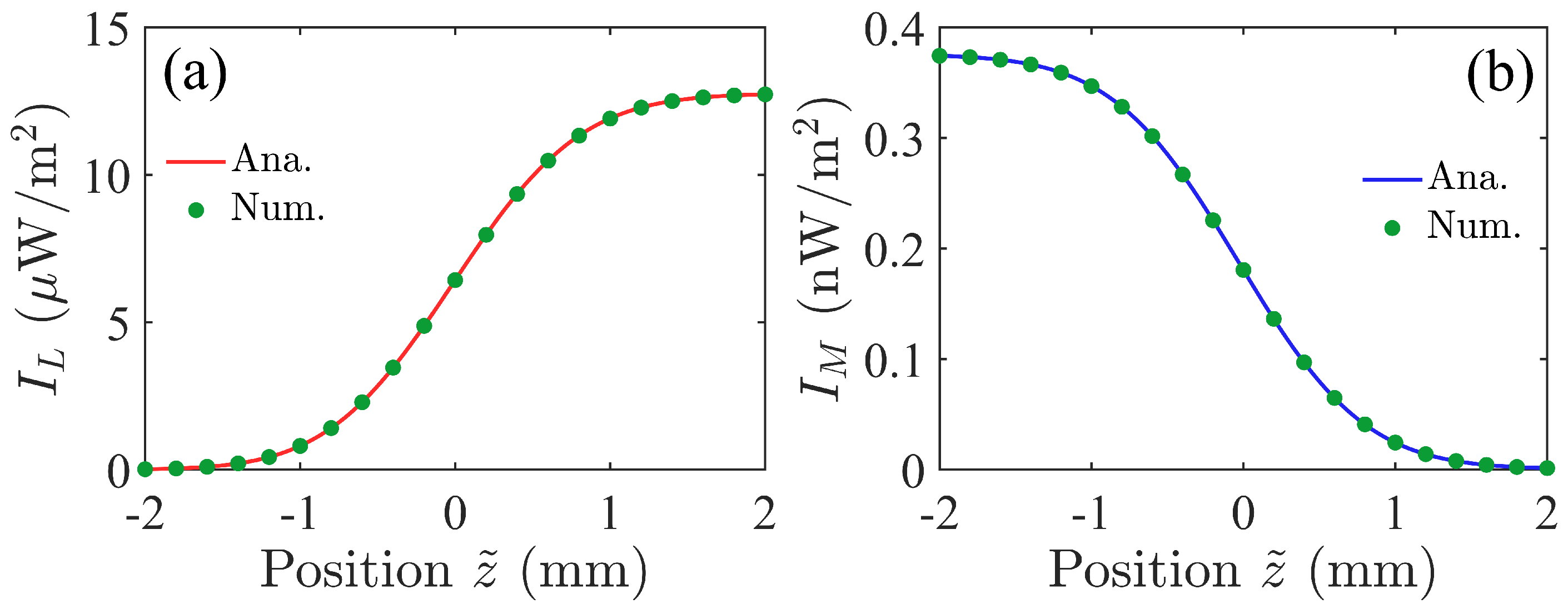}
	\caption{Analytical and numerical results showing the intensities of the converted field L in (a) and microwave field M in (b), versus position $\tilde{z}$. The simulation is performed with incident fields $\Omega_{P}=2\pi\times1$ kHz,  $\Omega_{M0}=2\pi\times0.01$ MHz, $\Omega_{C}=2\pi\times34$ MHz, $\Delta_{P}=\Delta_{M}=\Delta_{C}=0$, $\gamma=2\pi \times 10$ kHz, and $\Gamma=2\pi\times5.22$ MHz, and based on a Gaussian-distributed atomic cloud of peak atomic density $n_{at}=2.3\times10^{11}$ cm$^{-3}$ with $1/e^2$ radius $w_z=1$ mm (OD=99.9). The abbreviations "Ana." and "Num." stand for analytical and numerical results.}
	\label{f2}
\end{figure}

\begin{figure}[htbp]
	\centering
	\includegraphics[width=1.0\linewidth]{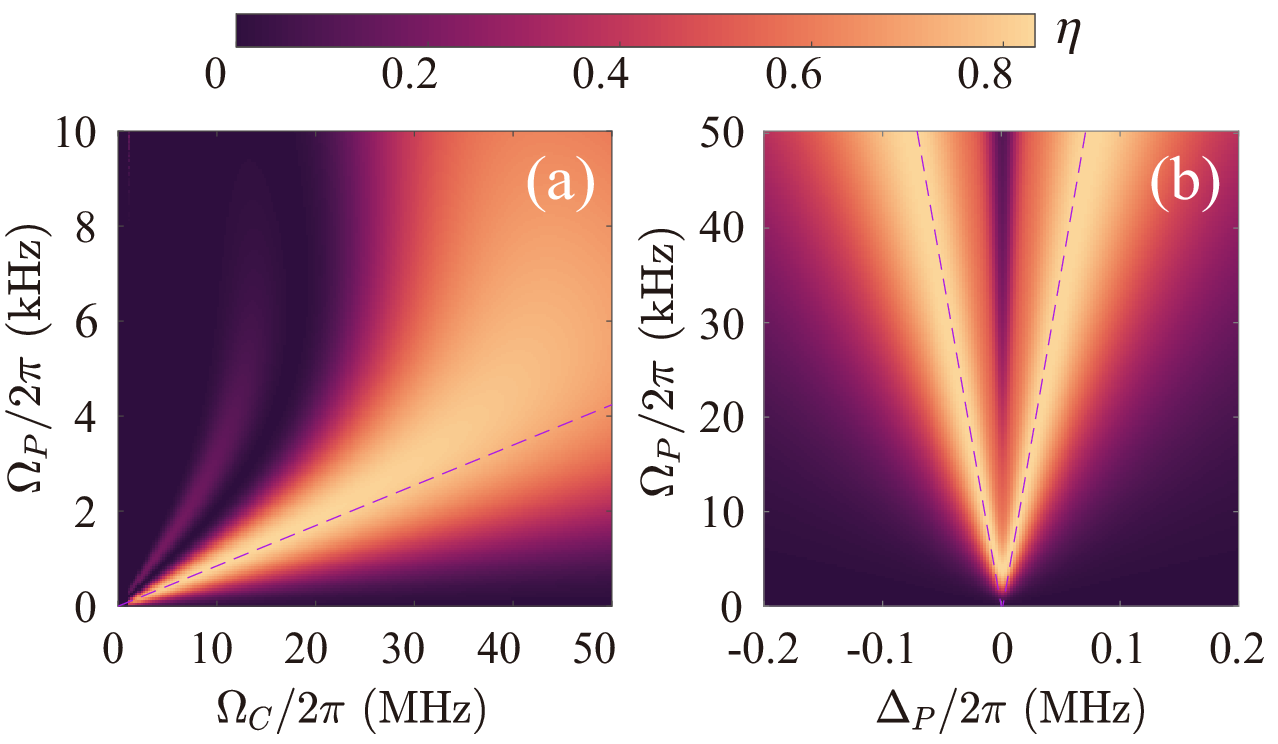}
	\caption{Numerical results of the conversion efficiency. (a) 2D density map of the efficiency for the all-resonant case versus $\Omega_{P}$  and $\Omega_{C}$. (b) 2D density map of the efficiency as a function of $\Omega_{P}$ and $\Delta_{P}$ computed with $\Omega_{C}=2\pi\times34$ MHz. The dashed line stands for the analytical results of Eq.~(\ref{eq10d}) in (a) and Eq.~(\ref{eq13}) in (b).  The other input parameters of the numerical simulation are the same as in Fig. \ref{f2},  except for OD=28.5.}
	\label{f3}
\end{figure}

We compute the stationary numerical solution of the Maxwell-Bloch equations \eqref{eq2} and \eqref{eq5a}-\eqref{eq5b} with the above parameters. 
Shown in Fig. \ref{f2} are typical spatial distributions of the L and M fields  intensities.  Our analytical model of Eqs. \eqref{eq6a}-\eqref{eq6b} exhibits outstanding agreement with the full-numerical simulation of the Maxwell-Bloch equations \eqref{eq2} and \eqref{eq5a}-\eqref{eq5b}.  Remarkably, Figs. \ref{f3}(a) and \ref{f3}(b) show that both all-resonant and off-resonant frequency-mixing configurations achieve near-unit photon conversion efficiencies. It should be noted that the parameters leading to efficient photon conversion are highly constrained, as observed in Fig. \ref{f3}(b). For example, given a detuning of the P field $\Delta_P$, there exists a well-defined Rabi frequency $\Omega_P$ that leads to maximum efficiency.
It should be also noted that the dark state condition of Eq. \eqref{eq8} depends on the amplitudes of the auxiliary fields, and is avoided by employing sufficiently large $\Omega_C$.
We provide a detailed analysis of these results in the next paragraphs, and investigate microwave-optical conversion along the lines of two possible scenarios: (1) all-resonant microwave-to-optical conversion, and (2) microwave-to-optical conversion where the auxiliary P and the M fields form a two-photon transition via off-resonant intermediate state. Finally, we attempt to formulate some principles for optimizing the efficiency.
\subsection{All-resonant microwave-to-optical conversion}
\label{acase}
As shown in the previous sub-section, near-unit photon conversion efficiency can be obtained on resonance when $\Omega_P< \gamma \ll \Omega_C$. Based on the result of section \ref{model}, it is possible to derive simplified expressions of the fields that provides a guidance to maximize the efficiency in the so-called strong parametric coupling field regime (\textrm{i.e.,} $\Omega_{C}\gg\Omega_{M}$,  $\Omega_{P}$) \cite{2010Jen}. For this, the density matrix $\rho$ is derived from perturbation theory assuming that the three fields, L, M, and P, are weak. From Eqs. \eqref{eq6a} and \eqref{eq6b}, and considering that $\Omega_{C}$ and $\Omega_{P}$ are positive real constants, the expressions for the L and M fields as well as the efficiency can be rewritten as follows:
\begin{subequations}
\begin{eqnarray}
\label{eq10a}
\Omega_{L}(\tilde{z})&\simeq&\Omega_{M0}\sqrt{\frac{\eta_{L}}{\eta_{M}}}\rm{Sin}\left(\tilde{z}\tilde{\beta}_{L}\sqrt{\eta_{M}\eta_{L}}\right),\\
\label{eq10b}
\Omega_{M}(\tilde{z})&\simeq&\Omega_{M0}\rm{Cos}\left(\tilde{z}\tilde{\beta}_{L}\sqrt{\eta_{M}\eta_{L}}\right),\\
\label{eq10c}
\eta&\simeq&\textrm{Sin}^{2}\left(\tilde{z}\tilde{\beta}_{L}\sqrt{\eta_{M}\eta_{L}}\right),
\end{eqnarray}
\end{subequations}
where $\tilde{\beta}_{L}$ is a constant that depends on $\Omega_{C}$, $\Omega_{P}$, $\gamma$, and $\Gamma$
\begin{eqnarray}
	\label{eq10e}
\tilde{\beta}_{L}=\frac{\Omega_{P}\Omega_{C}}{\gamma(\gamma\Gamma+\Omega_{C}^{2})}.
\end{eqnarray}
These expressions are obtained from a calculation of $\rho_{32}$ and $\rho_{41}$ by perturbation theory up to second order as shown in Appendix~A. As a result, the efficiency $\eta$ can reach 100\% when the following two conditions are met: $\Omega_{C}\gg\Omega_{P},\Omega_{M}$, and  
\begin{eqnarray}
	\label{eq10d}
	\tilde{z} \tilde{\beta}_{L}\sqrt{\eta_{M}\eta_{L}}=\pi/2.
\end{eqnarray}
This analytical expression displays good agreement with the numerical simulation as reported in Fig.~\ref{f3}(a).
\begin{figure}[htbp]
	\centering
     \includegraphics[width=1.0\linewidth]{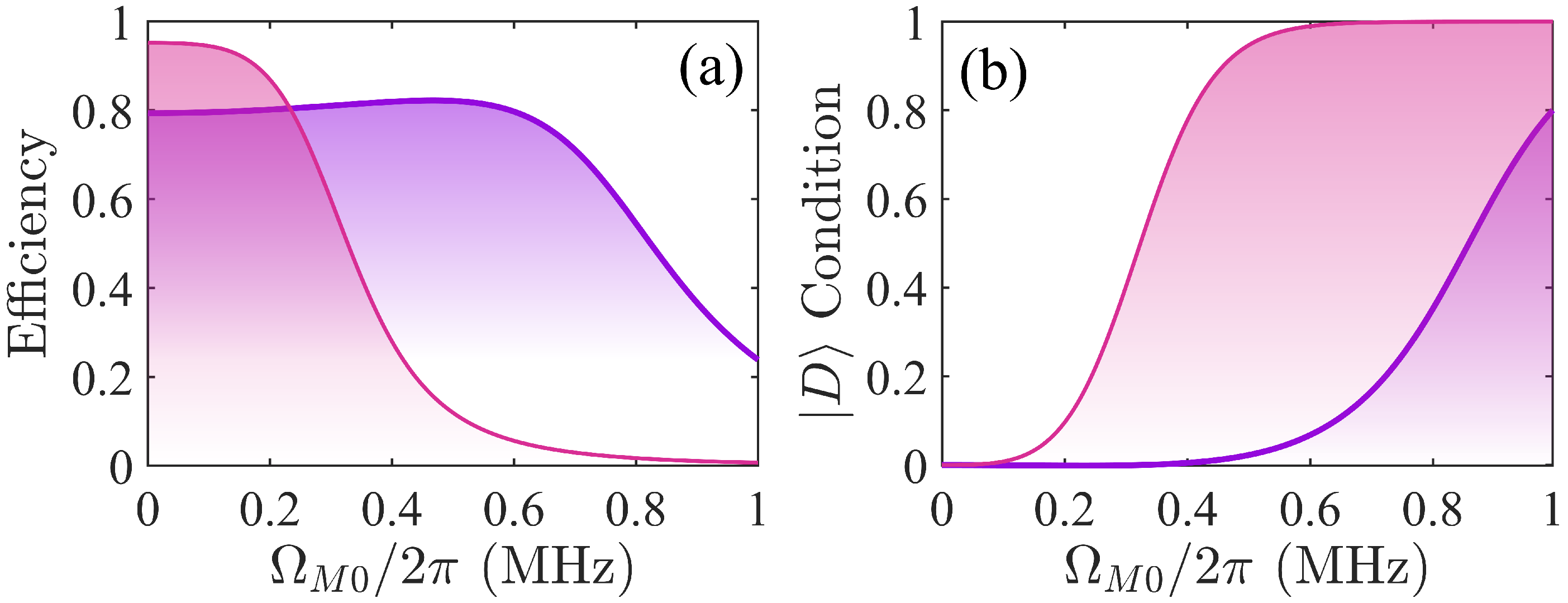}
	\caption{On-resonance frequency mixing: (a) conversion efficiency $\eta$; (b) dark state condition \eqref{eq8}. The numerical results are presented for two different $\Omega_{P}$ and as a function of $\Omega_{M0}$.  In (a) and (b), the thin pink lines are computed using $\Omega_{P}=2\pi\times1$ kHz with OD=99.9 and the thick purple lines using $\Omega_{P}=2\pi\times5$ kHz with OD=28.5, whereas the other parameters are similar to those in Fig. \ref{f2}, except for $\Omega_P$ and OD.}
	\label{f4}
\end{figure}

We perform full-numerical simulations of the efficiency of resonant four-wave mixing process as a function of the input microwave field $\Omega_{M0}$ to highlight possible effects of the dark states. The results are presented in Fig. \ref{f4}(a) for two different sets of input parameters.  With input auxiliary fields $\Omega_P=2 \pi \times 1$~kHz and $\Omega_C=2 \pi \times 34$~MHz, we  easily attain high conversion efficiency of more than 95\% at low microwave intensities when OD=99.9 [see pink thin line in Fig. \ref{f4}(a)]. While the transducer behaves linearly at small $\Omega_{M0}$, it becomes nonlinear once $\Omega_{M0}$ reaches a threshold of about $\Omega_{M0}^{\rm{Max}}=2 \pi \times 0.2$~MHz. 

The observed drop in efficiency for large $\Omega_{M0}$ is a result of population build up in the dark state $|D\rangle$ as the product $\Omega_{M}\Omega_L$ increases. The conversion comes rapidly to a halt due to decoupling of the atoms from all the fields once condition \eqref{eq8} is fulfilled. As shown in Fig. \ref{f4}(b), the dark state condition \eqref{eq8} reaches close to one during the conversion process with dark state probability of more that 99.99\% at the exit of the atomic cloud.

For widening the linear conversion range, one may increase $\Omega_{P}$ and $\Omega_{C}$ as suggested by the dark state condition \eqref{eq8}. 
For example,  with a five-fold larger $\Omega_P$ [thick purple line in Fig. \ref{f4}(a)], the transducer features a three-fold enhancement of the linear conversion range with efficiency as high as 80\%. Note that in the latter case, the required optical depth OD=28.5 is smaller, consistently with Eq.~\eqref{eq10d}.

\begin{figure}[htbp]
	\centering
	\includegraphics[width=1.0\linewidth]{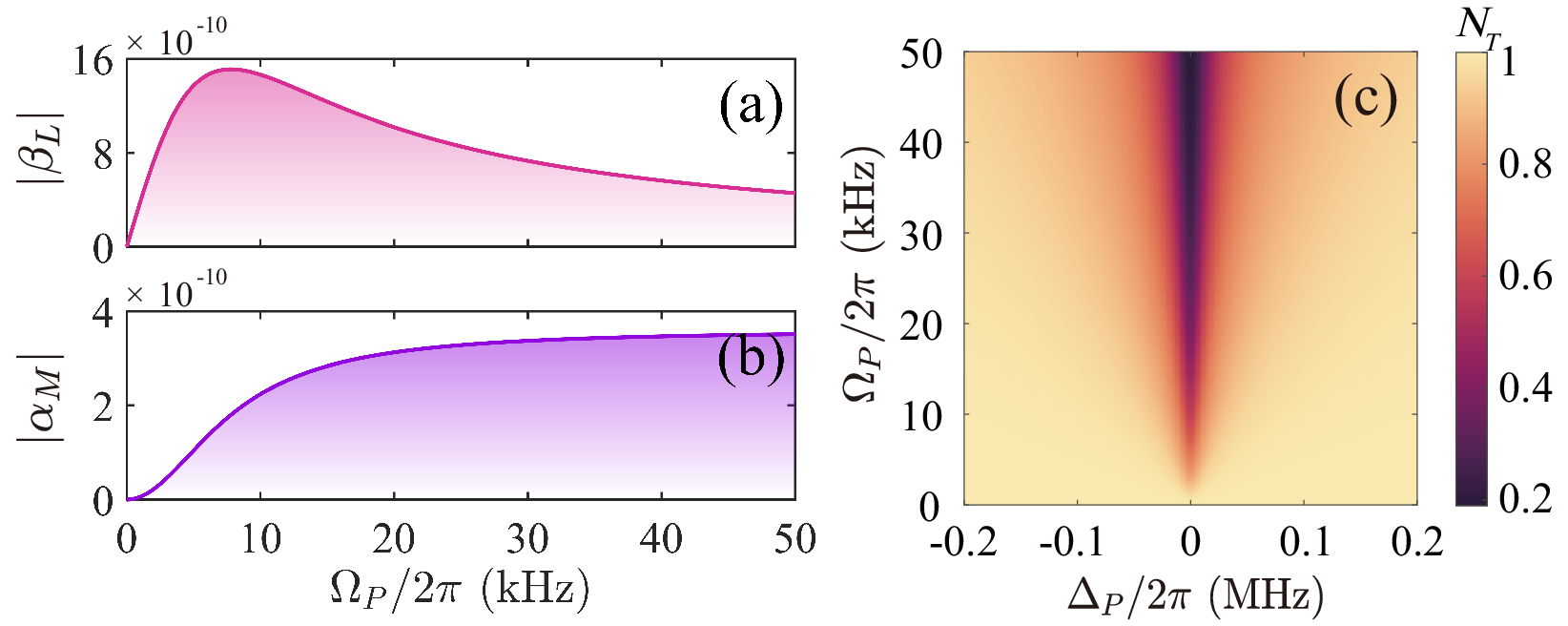}
	\caption{Microwave photon losses. (a) On-resonance parametric coefficient $|\beta_L|$ vs $\Omega_{P}$. (b) On-resonance self-coupling coefficient $|\alpha_M|$ vs $\Omega_{P}$. (c) Normalized total output photon flux $N_{T}$ vs $\Delta_{P}$ and $\Omega_{P}$.  The other input parameters of the numerical simulation are the same as in Fig. \ref{f2},  except for OD=28.5 and $\Omega_{C}=2\pi\times34$ MHz.}
	\label{f5}
\end{figure}

The observed reduction of the efficiency from 95\% to 80\% suggests that it may not be possible to augment $\Omega_{P}$ much further.
 Actually, when $\Omega_{P}\gg\gamma$, the bare atom states $\left | 1  \right \rangle$ and $\left | 2  \right \rangle$ are split into a doublet of dressed states $\psi_\pm$, as sketched in Fig. \ref{f1}(b). Most likely for this reason, the coupling of the atoms to the L field via parametric coefficient  $\beta_L(\approx\beta_M)$ is reduced at large $\Omega_P$, as confirmed by our numerical results in Fig. \ref{f5}(a). 
Moreover, the absorptive coefficient $\alpha_M$ increases with $\Omega_P$  as shown in Fig.~\ref{f5}(b). This occurs because, as the strength of the P field goes up, population in state $|2\rangle$ increases, and so does the absorption rate of microwave photons. Atoms excited in $|3\rangle$, with absorption of microwave photons, and then $|4\rangle$, do not couple well to the L field, hence subsequently decay back to state $|1\rangle$ via spontaneous emission, resulting in microwave photon losses, which is the main reason for the wide losses of the total number of photons in the L and M fields observed in Fig.~\ref{f5}(c). In that figure, we simulate the normalized total output photon flux $N_T$ of the M and L fields, given as $N_T=N_{total}/N_{0}=\eta+|\Omega_{M}(\tilde{z})/\Omega_{M0}|^2$, where $N_{total}$ is the total output photon flux, and $N_{0}$ is the input microwave photon flux. These photons are lost by spontaneous emission from state $|4\rangle $ after microwave photon absorption, and the losses are not due to absorption of the L field, since $\alpha_L$ is relatively constant versus $\Delta_P$ for the parameters of Fig.~\ref{f5}. 
Finally, both dressing by the P field and its adverse effect, microwave-induced fluorescence, are responsible for the reduced efficiency at large $\Omega_P$. We could decide to increase $\Omega_C$ instead of $\Omega_P$ for getting a larger dynamical range of the converter.  Employing larger $\Omega_{C}$ also helps in curbing down scattering as it leads to smaller population in state $|4\rangle$ considering the $|2\rangle\rightarrow |3\rangle \rightarrow |4\rangle$ $\Lambda$-system. However, using higher C field power comes at the price of needing larger OD in Eq.~\eqref{eq10d} for achieving high efficiencies, which may not be practical in experiments.

%
	%In this scenario, the ac-Stark splitting induced by the pump field P and control field C changes the resonance absorption conditions for the microwave M field and converted L field. This results in decoherence in the microwave field, and incurs additional absorption losses for the conversion Lfield at resonance.}
%

The discussion in this subsection explicitly demonstrates that careful selection of parameters based on dark state conditions and dressing by the auxiliary field P can always permit near-unit conversion efficiency in the regime of low microwave power.
Nonetheless, all-resonant four-wave-mixing poses challenges for photon conversion at larger microwave powers. Fortunately, the utilization of off-resonance wave mixing can circumvent these parameter limitations.
\subsection{Off-resonant microwave-to-optical conversion}
\label{bcase}
In the off-resonant scenario, we set the detuning $\Delta_{P}$ of the P field to be non-zero, while adjusting the detuning of the M field to maintain two-photon resonance, \textrm{i.e.,} $\Delta_{P}+\Delta_{M}=0$. As shown in sub-section \ref{expamle}, near-unit conversion efficiency is easily reachable for a given detuning provided that the correct Rabi frequency $\Omega_{P}$ is employed.

Let us again infer the conditions that provide maximum efficiency based on the perturbation theory results of  section \ref{model}. This is done without assuming that the P field is weak, since in the general case $\Omega_P$ is of the order of $\Delta_P$, and much larger than $\gamma$. Keeping only the dominant terms, the coherences $\rho_{32}$ and $\rho_{41}$ take the following simple form, after neglecting $\gamma$, and considering that $\Omega_P$ and $\Delta_P$ are much smaller than both $\Omega_C$ and $\Omega_C^2/\Gamma$:

\begin{subequations}
\begin{eqnarray}
\label{eq11a}
	\rho_{32}&\simeq&\frac{\Delta_{P}\Omega_{P}^*}{\Omega_{C}^*(2\Delta_{P}^2+|\Omega_{P}|^2)}\Omega_{L},\\
	\label{eq11b}
\rho_{41}&\simeq&\frac{\Delta_{P}\Omega_{P}}{\Omega_{C}(2\Delta_{P}^2+|\Omega_{P}|^2)}\Omega_{M}.
\end{eqnarray}
\end{subequations}
Substituting Eqs. \eqref{eq11a}-\eqref{eq11b} into Eqs. \eqref{eq5a}-\eqref{eq5b} and choosing $\Omega_P$ and $\Omega_C$ as real positive numbers, the analytical solutions for the L and M fields write as follows:
\begin{subequations}
	\begin{eqnarray}
\label{eq12a}
	\Omega_{L}(\tilde{z})&\simeq&i\Omega_{M0}\sqrt{\frac{\eta_{L}}{\eta_{M}}}\rm{Sin}\left(\frac{\sqrt{\eta_{L}\eta_{M}}\Omega_{P}\Delta_{P}}{\Omega_{C}(2\Delta_{P}^2+\Omega_{P}^2)}\tilde{z}\right),\\
	\label{eq12b}
\Omega_{M}(\tilde{z})&\simeq&\Omega_{M0}\rm{Cos}\left(\frac{\sqrt{\eta_{L}\eta_{M}}\Omega_{P}\Delta_{P}}{\Omega_{C}(2\Delta_{P}^2+\Omega_{P}^2)}\tilde{z}\right),\\
\label{eq12c}
\eta&\simeq&\textrm{Sin}^{2}\left(\frac{\sqrt{\eta_{L}\eta_{M}}\Omega_{P}\Delta_{P}}{\Omega_{C}(2\Delta_{P}^2+\Omega_{P}^2)}\tilde{z}\right).
\end{eqnarray}
\end{subequations}
Obviously, given the conditions set above, the efficiency $\eta$ may reach 100\% when

\begin{eqnarray}
\label{eq12}
\tilde{z}\sqrt{\eta_{L}\eta_{M}}\frac{\Omega_{P}\Delta_{P}}{\Omega_{C}(2\Delta_{P}^2+\Omega_{P}^2)}=\pm\pi/2.
\end{eqnarray}
It appears from Eq.~\eqref{eq12} that for a given set of parameters, there are four optimal values of the detuning $\Delta_{P}$. Numerically, the presence of $\gamma$ and its effect on absorption are not negligible for the smallest two solutions $|\Delta_P|$, and only the two largest optimal detunings $|\Delta_P|$ ensure that the efficiency $\eta$ is maximum. It is straightforward to obtain the expression for these detunings as

\begin{eqnarray}
\label{eq13}
\Delta_{P}=\pm\frac{\Omega_{P} \tilde{z}\sqrt{\eta_{L}\eta_{M}}+ \Omega_{P} \sqrt{\tilde{z}^2\eta_{L}\eta_{M}-2\pi^2 \Omega_{C}^2}}{2\pi \Omega_{C}}.
\end{eqnarray}
The expression~\eqref{eq13} of the optimal frequency detuning provides an excellent guide for obtaining large efficiencies, as reported on the density map in Figure \ref{f3}(b). For each $\Omega_P$, there exist two corresponding best detunings $\Delta_P$.

\begin{figure}[htbp]
	\centering
	\includegraphics[width=1.0\linewidth]{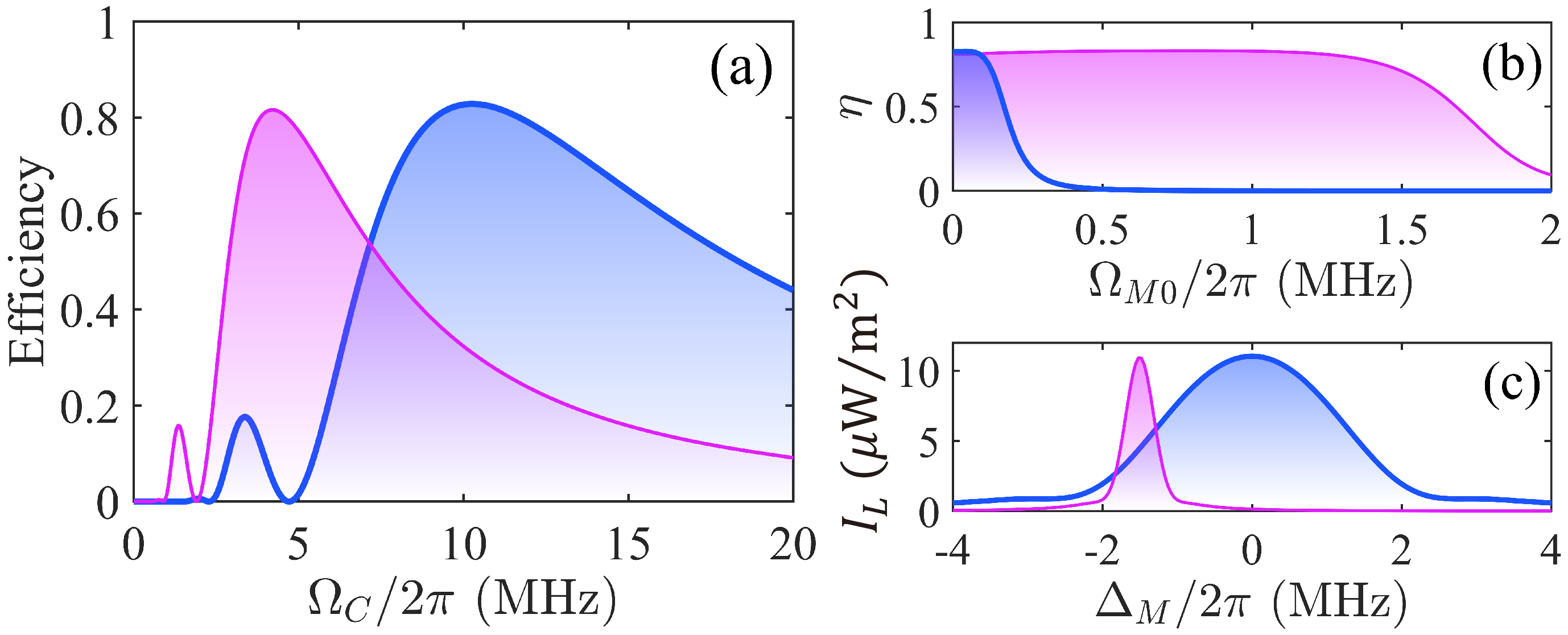}
	\caption{Comparison between on and off-resonance configurations. (a) Efficiency versus $\Omega_{C}$:  on-resonance case with $\Omega_{P}=2\pi\times1$ kHz (thick blue line); off-resonance case with $\Omega_{P}=2\pi\times0.1$ MHz and $\Delta_{P}=2\pi\times1.5$ MHz (thin magenta  line). In both cases, OD=28.5, and the other parameters are the same as in Fig. \ref{f2}. (b) Efficiency $\eta$ versus $\Omega_{M0}$, and (c) Intensity spectrum of the field L versus $\Delta_{M}$: all-resonant case with $\Omega_{C}=2\pi\times10$ MHz (thick blue line); off-resonant case with $\Omega_{C}=2\pi\times4$ MHz (thin magenta line). The other input parameters are the same as in (a).}
	\label{f6}
\end{figure}

The first advantage of the off-resonance configuration is to allow for using smaller $\Omega_C$.  As an example, comparing on and off-resonance configurations as a function of $\Omega_C$ for OD=28.5 in Fig. \ref{f6}(a), maximum efficiency is obtained at $\Omega_C=2 \pi \times 4$~MHz for the off-resonance case versus $2\pi \times 10$~MHz for on-resonance. Smaller $\Omega_{C}$ can be employed simply because fluorescence from state $|4\rangle$ following microwave photon absorption is largely absent with the off-resonance configuration (see Fig.~\ref{f5}). Too low $\Omega_{C}$ cannot be used though because large $\Omega_C$ also prevents direct scattering of the L field photons.  The oscillations of the efficiency in Fig. \ref{f6}(a) reach high values only at large enough $\Omega_{C}$ where EIT on the $|1\rangle\rightarrow |4\rangle \rightarrow |3\rangle$ ladder-scheme is efficient. 

The second advantage of the off-resonance configuration is that much larger Rabi frequencies $\Omega_P\sim\Delta_P$  are necessary when $\Delta_P\neq 0$. Hence, owing to the dark state condition \eqref{eq8}, the converter remains linear on a broad range of Rabi frequencies $\Omega_{M0}$, for example up to $2\pi\times1.5$ MHz in Fig.~\ref{f6}(b) using a detuning of $\Delta_{P}=2\pi\times1.5$~MHz.

A slight drawback of the off-resonance configuration is its reduced conversion bandwidth.
 Fig.~\ref{f6}(c) shows the normalized L field intensity as a function of the microwave frequency shift $\Delta_{M}-\Delta_{M0}$, where $\Delta_{M0}=-\Delta_P$  is the detuning at two-photon resonance. The spectrum is Lorentzian-shaped and centered around 0. Under the condition $\Delta_{M0}=0$, the full width at half maximum in Fig.~\ref{f6}(c) is approximately equal to $2\pi~\times$ 2.64~MHz, in comparison to about $2\pi~\times$ 0.45~MHz with detuning. This controllable and large bandwidth is one of the distinguishing features of the conversion in Rydberg atoms, and essential for extending the conversion scheme to the single-photon level.

\subsection{Optimization of the efficiency}
\begin{figure}[htbp]
	\centering
	\includegraphics[width=1.0\linewidth]{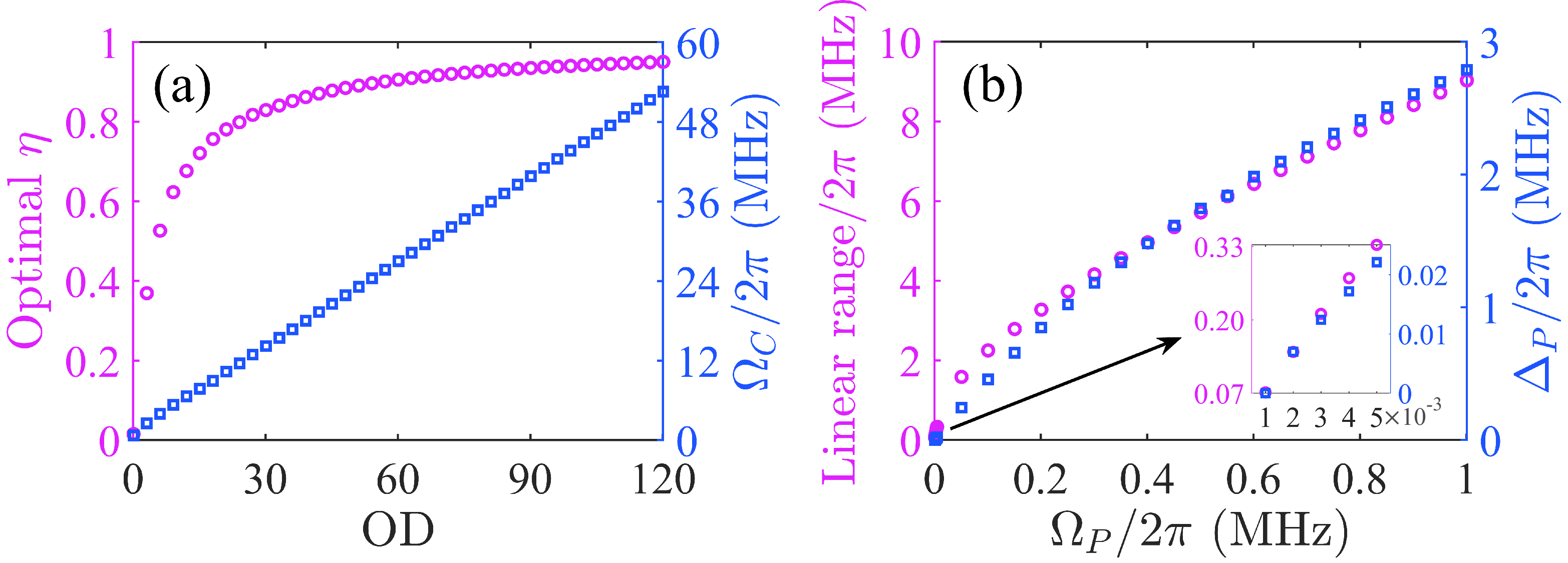}
	\caption{Full-numerical optimization. (a) Maximum efficiency (empty circles) and optimum $\Omega_{C}$ (empty squares) plotted versus OD for an optimization with variable parameters $\Delta_{P}$, $\Omega_{P}$ and $\Omega_{C}$. (b) Linear conversion range (empty circles) and optimum $\Delta_P$ (empty squares) versus $\Omega_P$ for an optimization with variable parameters $\Delta_{P}$ and $\Omega_{C}$, while keeping OD=28.5. The other input parameters are the same as in Fig. \ref{f2}.}
	\label{f7}
\end{figure}

At last, we investigate the conditions that lead to maximum efficiency through multivariate numerical optimization. Figure~\ref{f7}(a) presents the results of optimizing the efficiency $\eta$ with $\Omega_C$, $\Omega_P$, and $\Delta_P$ as free parameters, while ensuring $\Delta_P+\Delta_M=0$ and keeping $\Delta_C=0$. These last two conditions impose that the efficiency $\eta$ is a symmetric function of $\Delta_P$, hence we restrict the optimization to the blue detuning side, i.e. $\Delta_P>0$.
The optimal efficiency $\eta_{Max}$ increases rapidly with OD, is already of more than 80\% at OD $\approx 24.5$, and reaches 100\% asymptotically. The corresponding optimal value of $\Omega_{C}$ varies quasi-linearly with OD [see Fig.~\ref{f7}(a)], and as a result the ratio of optimized parameters $\frac{\Omega_{P}\Delta_{P}}{2\Delta_{P}^2+\Omega_{P}^2}$ is almost constant, consistently with Eq.~\eqref{eq12}. Relaxing the condition $\Delta_P+\Delta_M=0$ and adding $\Delta_M$ and $\Delta_C$ as free parameters, we have verified that the best results are still obtained for $\Delta_P+\Delta_M=0$ and $\Delta_C=0$. \\
\indent Moreover, the optimized efficiency depends little on the choice of $\Omega_P$. For instance, optimizing $\eta$ for OD = 28.5, and with $\Omega_C$ and $\Delta_P$ as free parameters, we find that $\eta_{Max}=82\pm 1\%$, hence is almost independent of the initial choice of $\Omega_P$. The one-to-one relationship between the optimized blue detuning $\Delta_P$ and $\Omega_P$ shown in Fig.~\ref{f7}(b) justifies that at low microwave power, any blue detuning $\Delta_P$ can be chosen to achieve the highest possible efficiency. A linear fit to Fig.~\ref{f7}(b) for $\Omega_P>2\pi\times0.4$~MHz yields  $\Delta_{P}\approx 2.18\ \Omega_{P}+  2\pi \times 0.65$~MHz.\\
\indent As explained in the two previous subsections, the reason for preferring the off-resonance configuration is to obtain an enhanced dynamical range. Defining the dynamical linear range as the maximum input microwave Rabi frequency $\Omega_{M0}^{\rm{Max}}$ for which $\eta_{Max}-\eta(\Omega_{M0}^{\rm{Max}})\le0.01$, we find that it rises approximately linearly as a function of $\Omega_{P}$, with two distinct slopes for $\Omega_P\lesssim \gamma$ and $\Omega_P\gtrsim\gamma$ [see circles in Fig. \ref{f7}(b)]. 
Linear fits to the data yield the dynamical linear range as $a\ \Omega_{P} + b$ ($a=65.42$ and $b\approx 0$ for $\Omega_{P}\lesssim\gamma$; $a=7.57$ and $b=2 \pi \times 1.47 $~MHz for $\Omega_{P}\gtrsim\gamma$). 

\section{Conclusions}\label{conc}
In conclusion, we have theoretically investigated microwave-to-optical conversion using collinear four-wave mixing via Rydberg atoms. We have demonstrated that photon conversion efficiencies exceed easily 80\% with large transducer's bandwidth $\sim 2 \pi \times 1.0$ MHz for applications relevant to quantum frequency conversion. In the all-resonant case, this is made possible by appropriately selecting parameters that avoid one of the dark states as well as microwave-induced fluorescence. The presence of the dark state is problematic at large microwave powers and in principle could be circumvented by simultaneously increasing the coupling field power and optical depth. Meanwhile, off-resonant frequency mixing allows efficient photon conversion to occur with a wider range of input microwave powers and with reduced power of the coupling laser field.

Our interpretations offer a fresh perspective on the conversion process compared to previous works, as we have been interested in optimizing the conversion over a wide range of parameters with a particular emphasis on the role of the dark states. The formalism was presented for four-wave mixing but also applies to six-wave mixing. Near-unit efficiencies can be achieved with both on-resonant and off-resonant six-wave mixing. Similarly to four-wave mixing, there is a dark state in the six-wave mixing system that may limit the dynamical range of the conversion, and this dark state is unchanged whether or not the fields are detuned from the intermediate states if the conditions of two-photon resonances are maintained. Similarly, the off-resonance configuration requires less power of the auxiliary optical fields. In general, four-wave mixing is more interesting as the required OD and total optical power are reduced.

Our research work was carried out under the assumption of an established steady state. Time-dependent calculations necessary for studying conversion with single photons may draw different conclusions~\cite{2019Petrosyan}. However, we can safely assume that microwave-induced fluorescence would remain an important issue of the all-resonant four-wave mixing configuration.

\section*{Acknowledgment}
This work is supported by the National Natural Science Foundation of China under grant number 12174460.
\section*{APPENDIX A}
\label{AppendixA}
\setcounter{equation}{0}
\renewcommand{\theequation}{A\arabic{equation}}
In this appendix, we derive the analytical expressions of the atomic coherences $\rho_{32}$ and $\rho_{41}$ provided in Eq. (\ref{eq4}) using perturbation theory \cite{2019Efficient}. The theory is presented in the case where the fields $\Omega_{M}$ and $\Omega_{L}$ are assumed to be sufficiently weak. A very similar theory is used in the second part of this appendix for the scenario where the P field is also considered as weak.

We assume that $\Delta_{P}+\Delta_{M}=0, ~\Delta_{C}=0$, and that the Rydberg state decay rates are equal $\gamma_2=\gamma_3=\gamma$. We expand the steady-state density matrix $\rho$ in perturbation series of the form
\begin{eqnarray}
	\label{A1}
\rho=\sum_{k=0}^{\infty} \rho^{(k)},
\end{eqnarray}
where $\rho^{(k)}$ denotes the contribution to $\rho$ in $k$-th order. The master equation Eq. (\ref{eq2}) may be rewritten as 
\begin{eqnarray}
\label{A2}
\partial_t\rho=\mathcal{L}_0 \rho+\mathcal{L}_1 \rho+\mathcal{L}_\Gamma \rho,
\end{eqnarray}
where
\begin{subequations}
\begin{eqnarray}
\label{A3a}
	\mathcal{L}_0 \rho=\frac{1}{2}\left[-i(\Omega_{P}A_{21}+\Omega_{C}A_{34})+\mathrm{H.c.}, ~~\rho\right],\\
\label{A3b}
	\mathcal{L}_1 \rho=\frac{1}{2}\left[-i(\Omega_{M}A_{32}+\Omega_{L}A_{41})+\mathrm{H.c.},~~\rho\right].
\end{eqnarray}
\end{subequations}
Note that we have neglected $\mathcal{L}_D \rho$ in Eq.~\eqref{A2}, assuming $\gamma_P,\gamma_C\ll \gamma$.
The zeroth-other solution in steady state, $\rho^{(0)}$ is obtained after keeping only the zeroth-order terms in Eq. \eqref{A2}, \textrm{i.e.,} solving the equation 
\begin{eqnarray}
\label{A4}
	\mathcal{L}_0 \rho^{(0)}+\mathcal{L}_\Gamma \rho^{(0)}=0.
\end{eqnarray}
Inserting the expansion of Eq. (\ref{A1}) into Eq. (\ref{A2}) leads in  steady state to a set of differential equations relating the $k$-th order solution $\rho^{(k)}$ to $\rho^{(k-1)}$ as 
\begin{eqnarray}
\label{A5}
	\mathcal{L}_0 \rho^{(k)}+\mathcal{L}_\Gamma \rho^{(k)}+\mathcal{L}_1\rho^{(k-1)}=0.
\end{eqnarray}
Starting from the result of Eq. \eqref{A4}, Eq. \eqref{A5} can be solved iteratively to yield all the orders $\rho^{(k)},~k>0$ under the constraints $\rm{Tr}(\rho^{(0)})=1$ and $\rm{Tr}(\rho^{(k)})=0 ~(k>0)$. The atomic coherences and populations of interest are given below to leading order:
 
\begin{subequations}
	\begin{eqnarray}
	\label{A4d}
	\rho_{32}^{(1)}&=& \alpha_{M}\Omega_{M}+\beta_{L}\Omega_{L},\\
	\label{A4e}
	\rho_{41}^{(1)}&=&\alpha_{L}\Omega_{L}+\beta_{M}\Omega_{M},\\
	\label{A4a}
	 \rho_{11}^{(0)}&=&1-\rho_{22}^{(0)},\\
	\label{A4b}
	\rho_{22}^{(0)}&=&\frac{|\Omega_{P}|^2}{\gamma^2+4\Delta_{P}^2+2|\Omega_{P}|^2},\\
	\label{A4c}
	\rho_{21}^{(0)}&=&-\frac{(i\gamma+2\Delta_{P}) \Omega_{P}}{\gamma^{2}+4\Delta_{P}^{2}+2|\Omega_{P}|^{2}},
	\end{eqnarray}
\end{subequations}
where 
\begin{subequations}
\begin{eqnarray}
\alpha_{M}&=&\frac{\rho_{22}^{(0)}\left(D_2(\gamma\Gamma+|\Omega_{C}|^2)-i\gamma|\Omega_{P}|^2\right)}{D_0}\nonumber\\
&&+\frac{\rho_{21}^{(0)}\Omega_{P}^*(|\Omega_{P}|^2-|\Omega_{C}|^2-i\Gamma D_2)}{D_0},\\
\beta_{L}&=&\frac{\rho_{11}^{(0)}(D_1\Omega_{C}\Omega_{P}^{*})+\rho_{12}^{(0)}\Omega_{C}(|\Omega_{P}|^{2}-|\Omega_{C}|^2-\gamma\Gamma)}{D_0},\\
\alpha_{L}&=&\frac{\rho_{11}^{(0)}\left(D_2(|\Omega_{P}|^2-2i\gamma D_3)-i\gamma|\Omega_{C}|^2\right)}{D_0}\nonumber\\
&&+\frac{\rho_{12}^{(0)}\Omega_{P}(|\Omega_{P}|^2-2i\gamma D_3-|\Omega_{C}|^2)}{D_0},\\
\beta_{M}&=&\frac{\rho_{22}^{(0)}D_1\Omega_{C}^*\Omega_{P}}{D_0}\nonumber\\
&&+\frac{\rho_{21}^{(0)}\Omega_{C}^*(|\Omega_{P}|^2-|\Omega_{C}|^2-2D_2D_3)}{D_0},\\
D_0&=&2\gamma^3\Gamma+|\Omega_{C}|^2\left(|\Omega_{C}|^2-D_4\right)-|\Omega_{P}|^2\left(iD_5+2|\Omega_{C}|^2\right)\nonumber\\
&&+|\Omega_{P}|^4+2\gamma^2\left(|\Omega_{C}|^2+|\Omega_{P}|^2-iD_6\right)\nonumber\\
&&-iD_7(3|\Omega_{C}|^2+|\Omega_{P}|^2-2i\Gamma\Delta_{P}).
\end{eqnarray}
\end{subequations}
and
\begin{subequations}
	\begin{eqnarray}
		D_1&=&-(2\Delta_{P}+2i\gamma+i\Gamma),\\
		D_2&=&-(2\Delta_{P}+i\gamma+i\Gamma),\\
		D_3&=&\Delta_{P}+i\gamma,\\
		D_4&=&\Delta_{P}(4\Delta_{P}+2i\Gamma),\\
		D_5&=&\Gamma(2\Delta_{P}+i\Gamma),\\
		D_6&=&\Gamma(3\Delta_{P}+i\Gamma),\\
		D_7&=&\gamma(2\Delta_{P}+i\Gamma).
	\end{eqnarray}
\end{subequations}
% %$D_1=i(2i\Delta_{P}-2\gamma-\Gamma),~D_2=i(2i\Delta_{P}-\gamma-\Gamma),~D_3=i\Delta_{P}-\gamma,~D_4=\Delta_{P}(4\Delta_{P}+2i\Gamma),~D_5=\Gamma(\Gamma-2i\Delta_{P}),~D_6=\Gamma(\Gamma-3i\Delta_{P}),~D_7=+\gamma(\Gamma-2i\Delta_{P})$.

%

Next, we apply similar perturbation theory in the strong parametric coupling regime, i.e. when the pump P field is also considered as a weak field. At zeroth order, $\rho_{11}^{(0)}= 1$ and $\rho_{22}^{(0)}=0$, in stark contrast with Eqs. \eqref{A4a} and \eqref{A4b}. It is obvious that this is only valid for $\Omega_P \ll \gamma$. In the strong parametric regime, the leading orders for $\rho_{32}$ and $\rho_{41}$ are
%$\alpha_{M}\approx\alpha_{M}\approx0$, and the conversion efficiency can be written as equation (\ref{eq8}).

\begin{subequations}
	\begin{eqnarray}
	\label{A5a}
	\rho_{32}^{(2)}&=&\frac{\Omega_{P}^{*}\Omega_{C}}{(2\Delta_{P}+i\gamma)(\gamma\Gamma+|\Omega_{C}|^{2})}\Omega_{L},\\
	\label{A5f}
	\rho_{41}^{(2)}&=&\frac{\Omega_{P}\Omega_{C}^{*}}{(2\Delta_{P}-i\gamma)(\gamma\Gamma+|\Omega_{C}|^{2})}\Omega_{M}.
	\end{eqnarray}
\end{subequations}

From these expressions we obtain $\alpha_{L(M)}=0$,
\begin{subequations}
	\begin{eqnarray}
		\label{A6a}
		\beta_{L}&=&\frac{\Omega_{P}^{*}\Omega_{C}}{(2\Delta_{P}+i\gamma)(\gamma\Gamma+|\Omega_{C}|^{2})},
	\end{eqnarray}
\end{subequations}
and $\beta_{M}=\beta_{L}^*$. Taking $\Omega_P$ and $\Omega_C$ as real constants, the leading orders of the coherences $\rho_{32}$ and $\rho_{41}$ become pure imaginary numbers in the all-resonant case.
Using Eq.~\eqref{A6a}, we can solve Maxwell-Bloch's equations (\ref{eq5a}-\ref{eq5b}) to obtain Eq. (\ref{eq10a}-\ref{eq10c}).

\section*{APPENDIX B}
\label{AppendixB}

In this appendix, we provide the theoretical framework for identifying all the dark states in our system.
\\
\indent A density matrix state $\rho_D$ is a dark state of the system if it has got no population in state $|4\rangle$, commutes with the Hamiltonian $H$, and verifies $\langle 3| \rho_D| 2\rangle =0$~\cite{2019Classification}. The first two conditions imply that $\rho_D=P \rho_D=\rho_D P$ and $[H, \rho_D]=0$, where $P$ is the projector on the subspace spanned by the three long-lived states $ |1\rangle$, $|2\rangle$, and $|3 \rangle $. We then obtain
\begin{eqnarray}
	\label{A6}
P[H, \rho_D]P=0 \Leftrightarrow [PHP, \rho_D]=0.
\end{eqnarray}
Therefore, $\rho_D$ and $PHP$ have got a common eigenbasis of three states $|\psi_k\rangle$ and $\rho_D$ may be written as  $\rho_D=\sum p_k |\psi_k\rangle \langle\psi_k|$, given $p_k\geq 0$ and $\sum p_k=1$. Because of $[H, \rho_D]=0$, $\rho_D$ must satisfy
\begin{eqnarray}
	\label{A7}
	\langle 4|[H, \rho_D]=0 \Leftrightarrow \sum p_k  \langle 4|H |\psi_k\rangle  \langle\psi_k|=0.
\end{eqnarray}
 As the $\langle\psi_k|$'s are independent vectors, this last identity is valid if and only if for every $k$, $4|H |\psi_k\rangle=0$. It is very restricting and implies that every $|\psi_k \rangle$ is proportional to a corresponding vector $|\varphi_k\rangle$ of the form
 \begin{eqnarray}
	\label{A8}
 	|\varphi_k \rangle=-\frac{\Omega_C^*}{\Omega_L}|1\rangle+\alpha_k |2\rangle +|3\rangle,
 \end{eqnarray}
where  $\alpha_k$ is a suitable coefficient. As $|\varphi_k \rangle$ is an eigenvector of $PHP$ with eigenvalue $\lambda_k$, the following system of equations must be satisfied
 \begin{eqnarray}
\label{A9}
\left\{
\begin{array}{ll}
	\frac{\Omega_P^*}{2}\alpha_k=-\lambda_k \frac{\Omega_C^*}{\Omega_L},\\
	\frac{\Omega_M}{2} \alpha_k -\delta=\lambda_k,\\
\frac{-\Omega_P}{2}\frac{\Omega_C^*}{\Omega_L}-\alpha_k\Delta_P+\frac{\Omega_M^*}{2}=\lambda_k \alpha_k,
\end{array}
\right.
 \end{eqnarray}
 where $\delta=\Delta_P+\Delta_M$.
The first two equalities in \eqref{A9} lead to 
 \begin{eqnarray}
	\label{A10}
	\left\{
	\begin{array}{ll}
		\alpha_k=\delta \frac{ 2 \Omega_C^*}{\Omega_P^*\Omega_L+\Omega_C^*\Omega_M}  &\ \textrm{if}\ \Omega_P^*\Omega_L \neq -\Omega_C^*\Omega_M\\
		\delta=0 &\ \textrm{if}\ \Omega_P^*\Omega_L = -\Omega_C^*\Omega_M.
	\end{array}
	\right.
\end{eqnarray}

Let us first consider the case $\Omega_P^*\Omega_L \neq -\Omega_C^*\Omega_M$ in Eq.~\eqref{A10}, which implies that the dark state, if it exists, is unique and a pure state proportional to $|\varphi_0\rangle$ of the form \eqref{A8} with the coefficient $\alpha_0$ obtained in Eq.~\eqref{A10}. Since $\langle 3| \rho_D| 2\rangle =0$, it comes directly that $\alpha_0=0$. From Eqs.~\eqref{A9}, it is obvious that this dark state exists if and only if $\delta=0$ and $\Omega_L\Omega_M^* = \Omega_P\Omega_C^*$, hence is equivalent to the one reported in Eq.~	\eqref{eq9}.

Let us now consider the second case in Eq.~\eqref{A10}, $\Omega_P^*\Omega_L = -\Omega_C^*\Omega_M$ and $\delta=0$. The third equation of the system \eqref{A9} possesses two solutions
\begin{eqnarray}
	\label{A11}
\lambda_{\pm}=\frac{-\Delta_P}{2}\pm \frac{1}{2}\sqrt{|\Omega_M|^2+|\Omega_P|^2+\Delta_P^2}.
\end{eqnarray}
Therefore, two possible states $|\varphi_k\rangle$, with $\alpha_k= 2 \lambda_{\pm}/\Omega_M$, are solutions to this system.
Denoting these two states as $|\varphi_{\pm}\rangle$, we then have $|\varphi_\pm \rangle=\frac{-\Omega_C^*}{\Omega_L}|1\rangle+2\frac{\lambda_\pm}{\Omega_M} |2\rangle +|3\rangle$, and

 \begin{eqnarray}
	\label{A12}
\langle 3| \rho_D |2 \rangle= p_+\frac{\langle \varphi_+|2\rangle}{|\langle \varphi_+|\varphi_+\rangle|^{1/2}} +p_-\frac{\langle \varphi_-|2\rangle}{|\langle \varphi_-|\varphi_-\rangle|^{1/2}}, 
\end{eqnarray}
with $p_\pm \geq 0 $, $p_+ + p_-=1$, and where we have used that $\langle 3|\varphi_{\pm}\rangle=1$.
From $\langle \varphi_+|\varphi_- \rangle=0$, and defining  $\langle \varphi_\pm|2\rangle=\alpha_\pm^*$, we can write
\begin{eqnarray}
 \label{A12a}
\alpha_+^*\alpha_-=-|\Omega_C/|\Omega_L|^2-1=-|\langle \varphi_\pm|\varphi_\pm\rangle|^2+\alpha_\pm^*\alpha_\pm.
\end{eqnarray}
Inserting this last result in Eq.~\eqref{A12}, we obtain 
 \begin{eqnarray}
	\label{A13}
	\langle 3| \rho_D |2 \rangle= \frac{p_- - p_+}{\alpha_- -\alpha_+},
\end{eqnarray}
such that $	\langle 3| \rho_D |2 \rangle=0$ when $p_\pm=1/2$. If the conditions $\Omega_P^*\Omega_L = -\Omega_C^*\Omega_M$ and $\delta=0$ are satisfied,  the dark mixed state of the system is 
 \begin{eqnarray}
 \label{A14}
 \rho_D=\frac{1}{2} \left(|+\rangle \langle +| +|-\rangle \langle -|\right),
 \end{eqnarray}
with  $|\pm \rangle=|\varphi_\pm\rangle/|\langle \varphi_\pm|\varphi_\pm\rangle|^{1/2}$.
In the limit of very small $\Omega_L$ and $\Omega_M$, $|\varphi_\pm\rangle$ are superpositions of states $|1\rangle$ and $|2\rangle$, and $\rho_D$ is simply equivalent to the steady state solution of a two-level atom interacting with the P field. When large $\Omega_P$ is employed, the state of the system is close to $\rho_D$,  featuring reduced coherence $\rho_{32}$, however, is a dark state only if $\Omega_P^*\Omega_L = -\Omega_C^*\Omega_M$. In practice, the system can only fall into the dark mixed state when $\gamma \ll \Omega_{P}$, as observed by inspection of the density matrix $\rho$ which then verifies Tr($\rho^2$)=1/2, $\langle +|\rho|+\rangle=1/2$, and $\langle -|\rho|-\rangle=1/2$. For $\Omega_P\ll\gamma$, the system tends to stay in the pure state $|1\rangle$, for which Tr($\rho^2$)=1 with $\langle +|\rho|+\rangle=1/2$ and $\langle -|\rho|-\rangle=1/2$.

In conclusion, two possible dark states may occur if and only if the two-photon detuning $\delta$ is set to zero, and their occurrence depends on the Rabi frequencies of the four fields. In practice, the dark mixed state is  not much observed in this system. It can limit the conversion for rates $\gamma$ at least one order of magnitude smaller than the one considered in our manuscript, and usually at sufficiently large $\Omega_P\sim\Omega_C$.
%
%
%\bibliographystyle{plain}
%\bibliography{Microwave_to_Optical_conversion_via_FWM_0314}
\end{document}